\documentclass[runningheads]{llncs}
\usepackage{graphicx}
\usepackage{multirow}
\usepackage[table,xcdraw]{xcolor}
\begin{document}
\title{Cranial Implant Design via Virtual Craniectomy with Shape Priors}
\author{Franco Matzkin\inst{1} \and 
Virginia Newcombe\inst{2} \and 
Ben Glocker\inst{3}\and 
Enzo Ferrante\inst{1}}

\institute{Research Institute for Signals, Systems and Computational Intelligence, sinc(i), CONICET, FICH-UNL (Argentina) \and 
Division of Anaesthesia, Department of Medicine, University of Cambridge (UK) \and 
BioMedIA, Imperial College London (UK)
}

\maketitle

\begin{abstract}
Cranial implant design is a challenging task, whose accuracy is crucial in the context of cranioplasty procedures. This task is usually performed manually by experts using computer-assisted design software. In this work, we propose and evaluate alternative automatic deep learning models for cranial implant reconstruction from CT images. The models are trained and evaluated using the database released by the AutoImplant challenge, and compared to a baseline implemented by the organizers. We employ a simulated virtual craniectomy to train our models using complete skulls, and compare two different approaches trained with this procedure. The first one is a direct estimation method based on the UNet architecture. The second method incorporates shape priors to increase the robustness when dealing with out-of-distribution implant shapes. Our direct estimation method outperforms the baselines provided by the organizers, while the model with shape priors shows superior performance when dealing with out-of-distribution cases. Overall, our methods show promising results in the difficult task of cranial implant design.


\end{abstract}

\section{Introduction}
Crainoplasty is a surgical procedure aimed at repairing a skull vault defect by insertion of a bone or nonbiological implant (e.g. metal or plastic) \cite{andrabi2017cranioplasty}. Such skull defect may exist due to different reasons, like a brain tumor removal procedure or a decompressive craniectomy surgery following a traumatic brain injury \cite{monteiro2020multiclass}. Cranial implant design is usually performed by experts using computer-aided design software specifically tailored for this task \cite{chen2017computer}. The AutoImplant challenge, organized for the first time at MICCAI 2020, aims at bench-marking the latest developments in computational methods for cranial implant reconstruction. In this work, we propose and evaluate two approaches to solve this task using deep learning models.

Previous works on skull and cranial implant reconstruction suggest that deep learning models are good candidates to solve this task. In \cite{morais2019automated} a denoising autoencoder was used to perform skull reconstruction, following an approach similar to the recently proposed Post-DAE method \cite{larrazabal2019,larrazabal2020}. In this case, a denoising autoencoder is trained to reconstruct full skulls from corrupted versions. However, the model proposed in \cite{morais2019automated} works with skulls extracted from magnetic resonance images, can only handle low resolution images and was evaluated on the full-skull reconstruction task. Here we focus on reconstructing the flap only, on skulls extracted from high resolution and anisotropic computed tomography (CT) images. Other approaches rely on a head symmetry assumption and propose to take advantage of it to reconstruct the missing parts by mirroring the complete side of the skull \cite{Hieu2003}. However, this is not a realistic assumption since missing flaps may occur in both  sides simultaneously. Another alternative could be the subtraction of the aligned pre- and post-operative CT scans. Unfortunately, this requires to have access to the pre-operative image, which may not be the case in real clinical scenarios. 

Recently, we have proposed \cite{matzkin2020} a simple virtual craniectomy procedure which enables training different deep learning models in a self-supervised way, given a dataset composed of full skulls. In this work, we compared two different approaches: direct estimation of the implant, or reconstruct-and-subtract (RS) strategies where the full skull is first reconstructed, and then the original image is subtracted from it to generate a difference map. We evaluated different architectures and concluded that direct estimation produces more accurate estimates than RS strategies, since the latter one tends to generate noise in areas far from the flap. A different approach has been introduced by the AutoImplant challenge organizers \cite{li2020baseline} which also employs deep learning models, but it works in two steps. First, a low resolution version of the image is reconstructed to localize the area where the defected region is located. Then, they extract a 3D patch from the high resolution image and process it using a second neural network trained for fine implant prediction. 

In this work, based on the conclusions from \cite{matzkin2020}, we employ a direct estimation method that operates on full skulls which are rigidly registered to an atlas and resampled to an intermediate resolution. Aligning the images allow us to work in a common space which simplifies the reconstruction task. We adapt the virtual craniectomy procedure to account for more realistic flap shapes, similar to the ones introduced in the AutoImplant challenge. Moreover, we propose to incorporate anatomical priors into the standard direct estimation model introduced in \cite{matzkin2020} by feeding the registered skull atlas as an extra image channel. Previous works \cite{lee2019tetris} have shown that incorporating approximate shape priors as additional image channels is a simple yet effective way to increase the anatomical plausibility of the segmentations, since it provides supplementary context information to the network. We compare the results of our two methods with those obtained by the baseline benchmark model introduced in \cite{li2020baseline}, showing the superiority of our approach.

\begin{figure}[t]
    \centering
    \includegraphics[width=.95\linewidth]{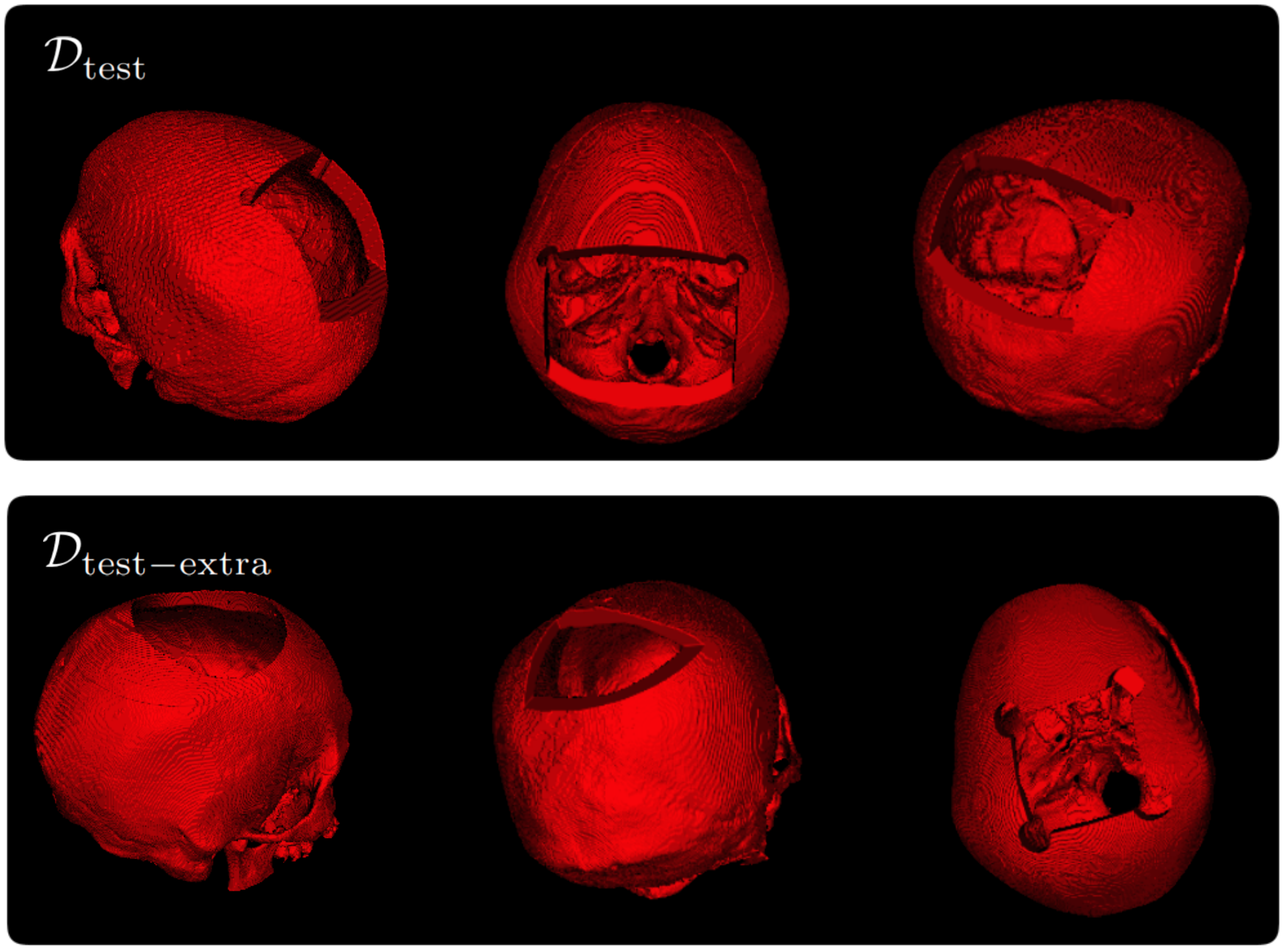}
    \caption{Examples of images from the $\mathcal{D}_\mathrm{test}$ set and $\mathcal{D}_\mathrm{test-extra}$ (out-of-distribution cases). As it can be observed, images from $\mathcal{D}_\mathrm{test}$ follow a common pattern, while those in $\mathcal{D}_\mathrm{test-extra}$ present different defects with various shapes.}
    \label{fig:test-images}
\end{figure}

\section{Challenge description and database}
The AutoImplant challenge organizers provided 100 images for training ($\mathcal{D}_\mathrm{train}$) and 110 images for testing. From the 110 test images, 100 of them (denoted here as $\mathcal{D}_\mathrm{test}$) have simulated surgical defects which follow the same distribution as the ones on the training images, while the remaining 10 (denoted as $\mathcal{D}_\mathrm{test-extra}$) have defects which do not follow the same distribution (see Figure \ref{fig:test-images}). The images were selected from the CQ500 public database\footnote{The database can be accessed at: \url{http://headctstudy.qure.ai/dataset}} \cite{chilamkurthy2018development}. They have fixed image dimension in the axial plane (512 x 512) and a variable number of axial slices Z.

\sloppy The training dataset ($\mathcal{D}_\mathrm{train}$) is composed of triplets ($\mathcal{X}^\mathrm{full}, \mathcal{X}^\mathrm{defected}, \mathcal{Y}$), where $\mathcal{X}^\mathrm{full}$ is the full skull, $\mathcal{X}^\mathrm{defected}$ corresponds to the defected skull and $\mathcal{Y}$ to the removed defect that we aim at reconstructing. For the test images, only the $\mathcal{X}^\mathrm{defected}$ images were released. We evaluated the proposed methods in the test images and submitted the results to the organizers, who computed the metrics reported in this paper. It is important to note that, in order to avoid overfitting to the test data, we could submit our results a maximum of 5 times.

\section{Methods}
\begin{figure}[t]
    \centering
    \includegraphics[width=\textwidth]{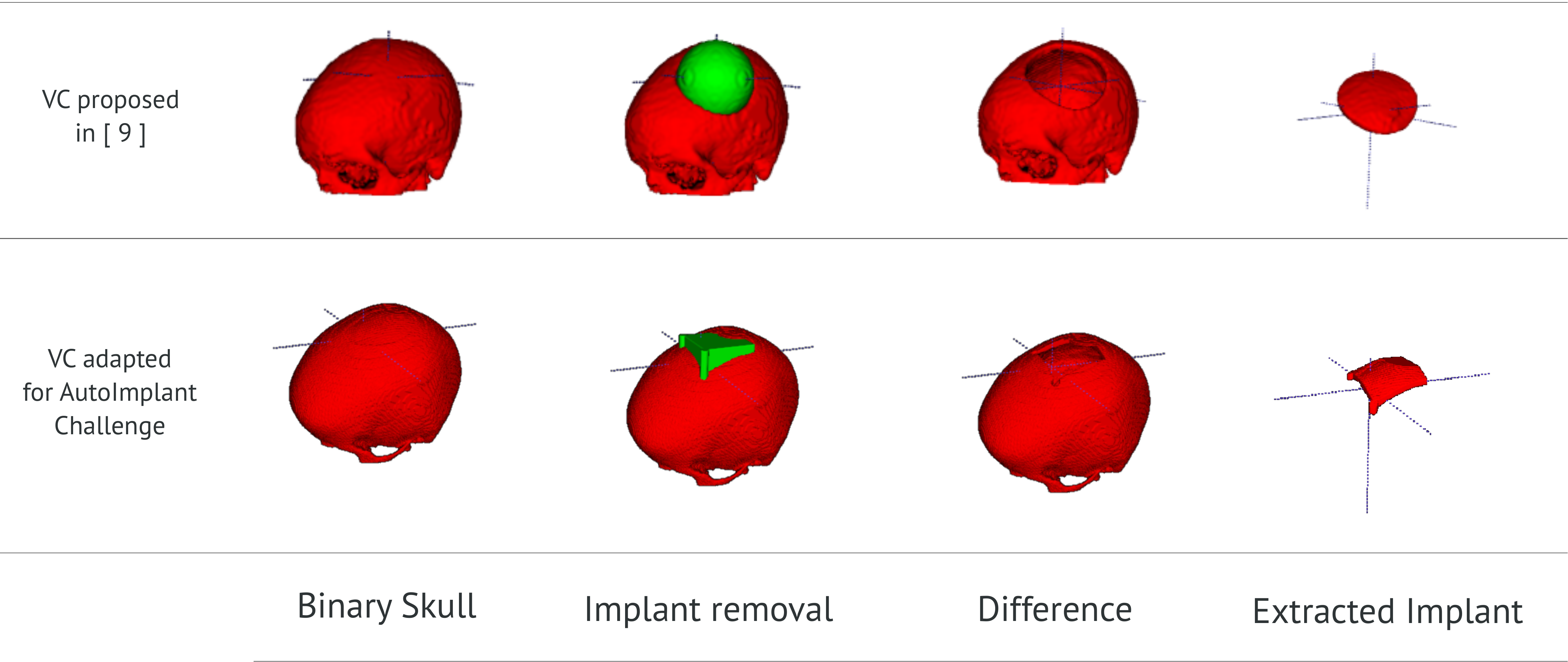}
    \caption{Modified virtual craniectomy procedure. We incorporated new template shapes for the virtual craniectomy to account for the pattern found in the AutoImplant challenge dataset.}
    \label{fig:virtualCraniectomy}
\end{figure}

The proposed cranial implant reconstruction methods operate on the space of binary volumetric masks. Such binary skull can be obtained by simply thresholding a brain CT image according to the Hounsfield scale, or applying more sophisticated methods. In the AutoImplant challenge, the skulls were already provided as binary volumes, extracted from the CT images using thresholding and additional post-processing steps (for further details we refer to \cite{li2020baseline}). Since the training data includes the full skulls, we leveraged the virtual craniectomy procedure proposed in \cite{matzkin2020} to train our models.

\subsection{Virtual Craniectomy and Data Augmentation}
Given a full skull, we designed a virtual craniectomy procedure which consists in removing a bone flap using a template located in a random position along its upper part. In \cite{matzkin2020}, spherical template shapes were used. By visual inspection of the AutoImplant training data, we observed that defects tend to follow a pattern given by the intersection of the skull with a cube with two cylinders over the edges perpendicular to the axial planes. So, we designed a variable-size template shape which produces similar defects, as shown in Figure \ref{fig:virtualCraniectomy}. To increase the diversity of our training procedure, we also included spherical and cubic templates of random sizes (all of the three shapes were selected with equal probability).

The virtual craniectomy was used as a data augmentation mechanism to generate a variety of training samples from a limited amount of full skulls, resulting in a self-supervised learning approach, where no annotated skull defects are required for training. We also included salt and pepper noise in the input images with probability 0.01. Moreover, we also considered the defective skulls provided by the organizers as part of our datasets (in these cases,  virtual craniectomy was not performed). During training, we sampled images coming from both sources: simulated virtual craniectomies and defective skulls provided by the organizers. 

\subsection{Common space alignment}
 Before training, all the images were rigidly registered to a common space determined by a full skull atlas. It consists in a thresholded version of a full-skull head CT atlas constructed by averaging several healthy head CT images. Such atlas allowed us to normalize the images by resampling them to an intermediate resolution. We chose this resolution to be 0.695 x 0.695 x 0.715 mm (resulting in a volume of 304 x 304 x 224 voxels) because it was the maximum size we managed to fit in GPU memory. Moreover, aligning the images in a common space simplifies the reconstruction task for the neural network, since it can focus on shape variations which are more relevant to the reconstruction task than translations and rotations. We used the FLIRT software package \cite{jenkinson2002improved} for rigid registration. At test time, given a test defective image $\mathcal{X}^\mathrm{defected}_i$, we apply the same registration procedure which returns a transformation $\mathcal{T}$ and its inverse $\mathcal{T}
^{-1}$. The transformation is applied to the original image $\mathcal{T} \circ \mathcal{X}^\mathrm{defected}_i$. The estimated skull defect $
\hat{\mathcal{Y}}_i$ is reconstructed in the common space, and the final estimate in the original space is recovered by applying the inverse transformation $\mathcal{T}^{-1} \circ \hat{\mathcal{Y}}_i$.
\begin{figure}[t]
    \centering
    \includegraphics[width=\textwidth]{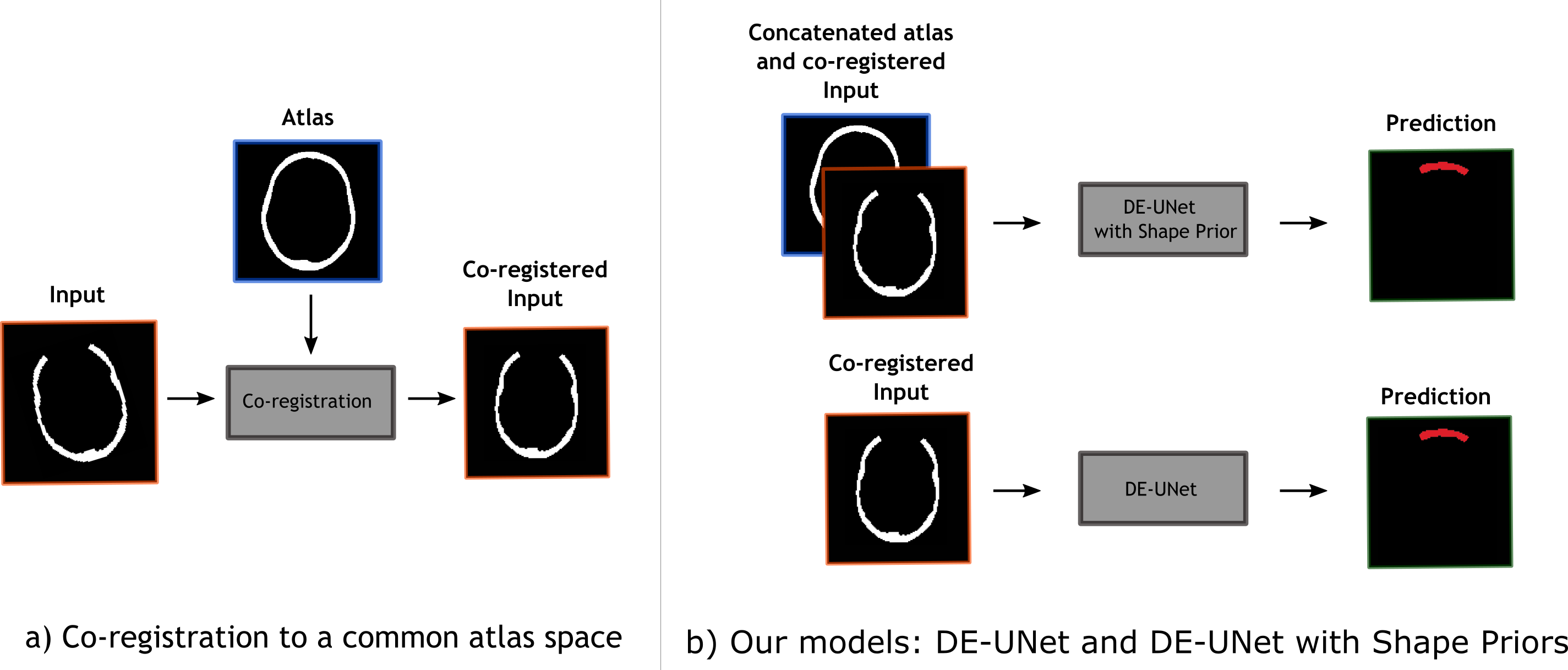}
    \caption{\textbf{(a)} The images are first registered to an atlas space, and resampled to a common resolution. We store the resulting transform $\mathcal{T}$ and its inverse $\mathcal{T}^{-1}$. \textbf{(b)} We compare two different approaches for the implant reconstruction task. The first one is a standard DE-UNet model. The second one incorporates a shape prior by considering the atlas as an extra input channel to the network. After prediction, the segmentation mask is mapped-back to the original image space using the inverse transform $\mathcal{T}^{-1}$.}
    \label{fig:models}
\end{figure}

\subsection{Direct estimation}
Our first method is a direct estimation model which follows the same architecture as the DE-UNet  used in \cite{matzkin2020}. It is a standard 3D UNet encoder-decoder architecture with skip connections, trained using a compound loss which combines Dice and cross-entropy terms \cite{Patravali18} (for more details, we refer to our work in \cite{matzkin2020}). After reconstruction, the segmentation is re-mapped to the original resolution using the inverse transform $\mathcal{T}^{-1}$ as previously discussed.

The model is trained using batches with full volume images, pre-aligned in the common space and resampled to an intermediate resolution as previously discussed.

\subsection{Direct estimation with shape priors}
Since the DE-UNet model is a fully convolutional architecture, the receptive field of the model is mainly determined by the amount of layers and parameters of the pooling and convolution operations. In other words, the local support of the output predictions is restricted to a certain area in the input image. When we have to reconstruct big or out-of-distribution skull defects, it may happen that most of the image support for certain parts of it are background, so the network may have no context to infer the implant shape. To overcome this limitation and make our model robust, we propose to incorporate context via shape priors given as an extra channel to the segmentation network. Previous works \cite{lee2019tetris} have shown that this simple extension can boost the robustness of existing state-of-the-art pixel-wise approaches in medical image segmentation tasks. 

We take advantage of the fact that images are co-registered to a common space, and use the same skull atlas as shape prior. After registration, we concatenate the resampled image with the atlas as a extra input channel, and train the network following the same strategy discussed before. In this case, the shape prior acts as a kind of initialization for the network's output, providing additional context that will be useful specially to reconstruct out-of-distribution defects. We refer to this model as DE-Shape-UNet.
\begin{figure}[t]
    \centering
    \includegraphics[width=\textwidth]{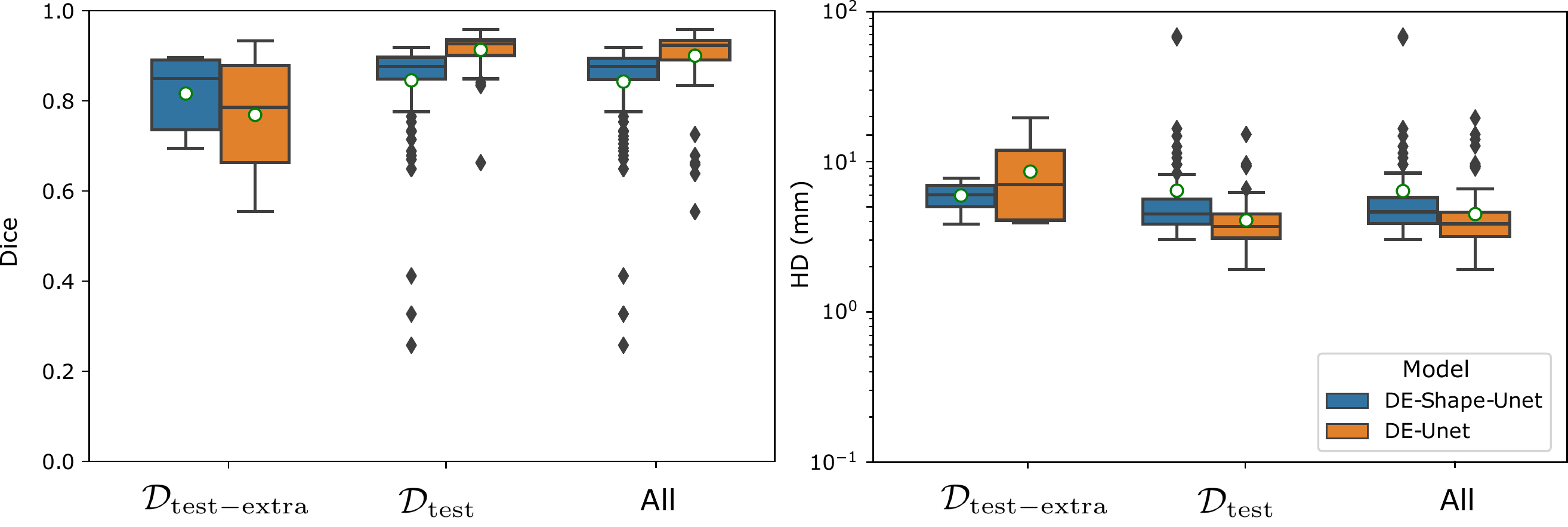}
    \caption{Comparison of the results for the proposed methods in terms of Dice and Hausdorff Distance (HD). HD is shown in log scale for better visualization.}
    \label{fig:quanitative}
\end{figure} 

\subsection{Implementation details}
The models were implemented in Pyhton, using the PyTorch 1.4 library. We trained and evaluated the CNNs using an NVIDIA TITAN Xp GPU with 12GB of RAM. The same virtual craniectomy and data augmentation procedure was used to train both models. In both cases we used a compound loss function which combines Dice loss and Binary Cross Entropy (BCE) as $L  =  L_{Dice} + \lambda L_{BCE}$ (parameter $\lambda$ was set to $\lambda = 1$ by grid search). Both models followed the DE-UNet architecture described in \cite{matzkin2020}; the only difference between them was that we concatenated the atlas as an extra input channel in the DE-Shape-UNet model. For optimization, we used Adam with initial learning rate of 1e-4. The batch-size was set to 1 for memory restrictions. The models were trained for 50 epochs. The 100 training images were split in 95 images for training and 5 for validation. After 50 epochs, we kept the model that achieved best accuracy in the validation fold.

\section{Results}
\begin{figure}[]
    \centering
    \includegraphics[width=\linewidth]{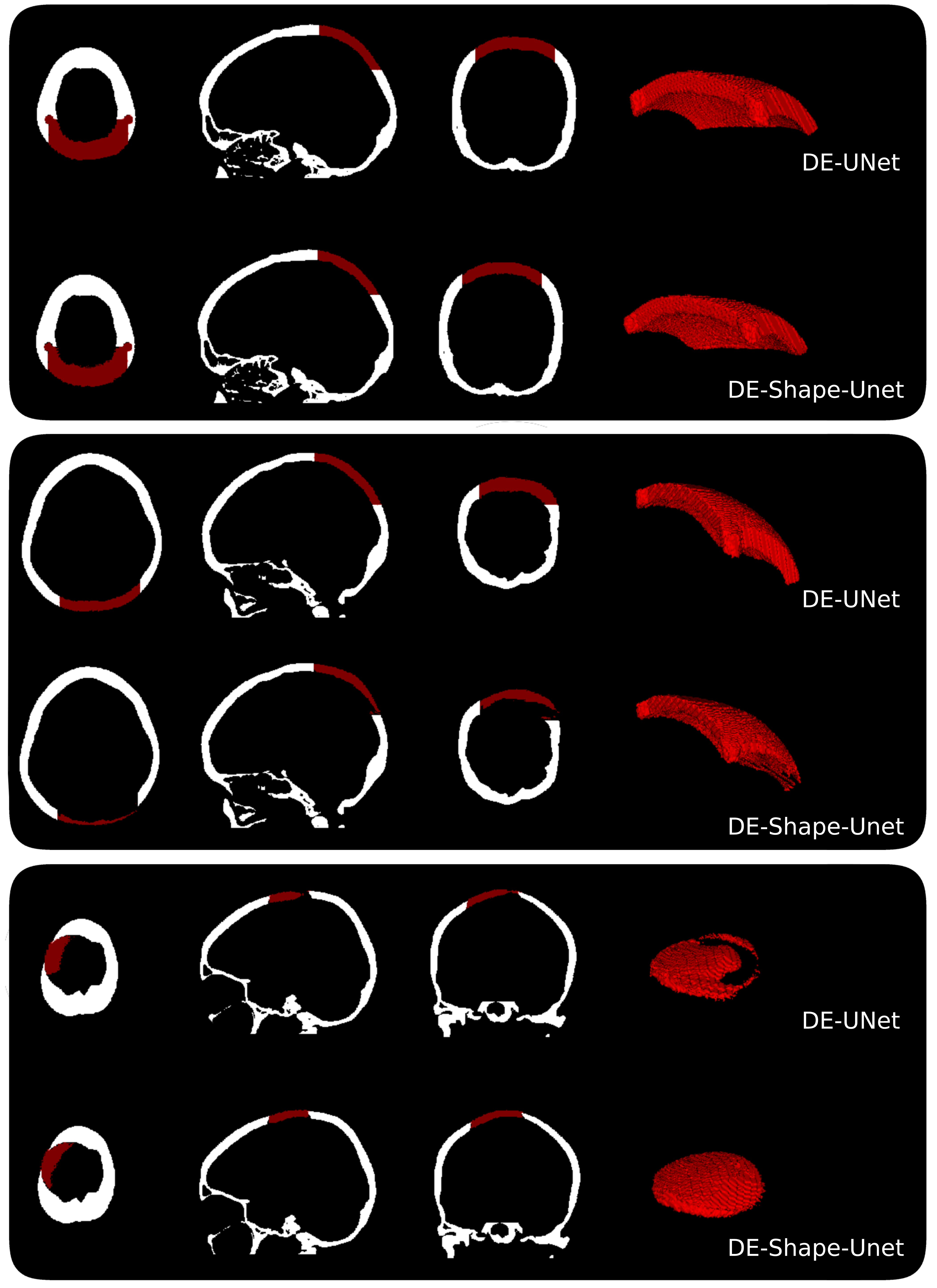}
    \caption{Examples of different reconstructions from $\mathcal{D}_\mathrm{test}$ (cases which follow the same pattern than the training dataset, shown in rows 1 and 2) and $\mathcal{D}_\mathrm{test-extra}$ (out-of-distribution case, shown in row 3). As we can observe, both methods performed well in the image depicted in row 1. For the case in the 2nd row, even if the DE-Shape-UNet model managed to reconstruct the implant, the quality of the reconstruction is lower than that of the DE-UNet. The opposite happened with the image in row 3 (an out-of-distribution case from $\mathcal{D}_\mathrm{test-extra}$) where the model which incorporated shape priors managed to reconstruct the implant, while the DE-UNet failed in this task.}
    \label{fig:qualitative}
\end{figure}

Figure \ref{fig:quanitative} and Table \ref{tab:my-table} include a quantitative comparison of the results. We report Dice coefficient and Hausdorff distance measured in the $\mathcal{D}_\mathrm{test}$ (100 images), $\mathcal{D}_\mathrm{test-extra}$ (10 images) and the whole test dataset. We observe that DE-Shape-UNet presents better performance for out-of-distribution cases ($\mathcal{D}_\mathrm{test-extra}$), while DE-UNet outperforms the other model in the $\mathcal{D}_\mathrm{test}$ set. Since the whole test dataset is composed of 100 images from $\mathcal{D}_\mathrm{test}$ and only 10 images from $\mathcal{D}_\mathrm{test-extra}$, the DE-UNet model shows better performance in the overall comparison. Moreover, DE-UNet model outperforms the two baseline models (N1 and N2) reported by the organizers in \cite{li2020baseline}. Figure \ref{fig:qualitative} provides some visual examples for reconstructions obtained with both methods in samples from $\mathcal{D}_\mathrm{test}$ and $\mathcal{D}_\mathrm{test-extra}$.

\begin{table}[t]
\centering
\caption{Quantitative results obtained for the two proposed methods (DE-UNet and DE-Shape-UNet) compared with the two baselines reported by the challenge organizers in \cite{li2020baseline}. We report the mean Dice and HD values, and the standard deviation in parentheses.}
\resizebox{\textwidth}{!}{\begin{tabular}{ccccccc}
\multirow{2}{*}{Method} & \multicolumn{2}{c}{$\mathcal{D}_\mathrm{test}$ (100)} & \multicolumn{2}{c}{$\mathcal{D}_\mathrm{test-extra}$ (10)} & \multicolumn{2}{c}{Overall} \\ \cline{2-7} 
 & Dice & HD (mm) & Dice & HD (mm) & Dice & HD (mm) \\ \hline
Baseline N1 \cite{li2020baseline} & 0.809 & 5.440 & - & - & - & - \\
Baseline N2 \cite{li2020baseline} & 0.855 & 5.182 & - & - & - & - \\
DE-UNet & \textbf{0.913 (0.038)} & \textbf{4.067 (1.762)} & 0.769 (0.126) & 8.585 (5.128)& \textbf{0.900 (0.067)} & \textbf{4.477 (2.626)} \\
\multicolumn{1}{l}{DE-Shape-UNet} & 0.845 (0.107) & 6.414 (9.060) & \textbf{0.816 (0.078)} & \textbf{5.952 (1.258)} & 0.842 (0.105) & 6.372 (8.648) \\ \hline
\end{tabular}}
\label{tab:my-table}
\end{table}

\section{Conclusions}
In this work, we evaluated two different approaches for cranial implant reconstruction based on deep learning: a direct estimation method and an alternative strategy which incorporates shape priors. We adapted the virtual craniectomy procedure proposed in \cite{matzkin2020} to the defect distribution of the AutoImplant challenge. We found that the simple DE-UNet method produces more accurate results for the skull defects which follow the same distribution as those in the training dataset. However, for out-of-distribution cases where the DE-UNet model tends to fail, the use of shape priors increases the robustness of the model, providing additional context to the network. In our implementation, this gain in robustness for out-of-distribution cases was achieved to the detriment of the overall accuracy. In future work, we plan to study alternative ways to introduce shape priors, e.g. considering deformable registration with anatomical constraints \cite{mansilla2020learning} to the atlas space instead of rigid transformations, or incorporating shape priors in a co-registration and segmentation process \cite{shakeri2016prior}.

\section{Acknowledgments}
The authors gratefully acknowledge NVIDIA Corporation with the donation of the Titan Xp GPU used
for this research, and the support of UNL (CAID-PIC-50220140100084LI) and
ANPCyT (PICT 2018-03907).
\bibliographystyle{splncs04}
\bibliography{references}

\begin{thebibliography}{10}
\providecommand{\url}[1]{\texttt{#1}}
\providecommand{\urlprefix}{URL }
\providecommand{\doi}[1]{https://doi.org/#1}

\bibitem{andrabi2017cranioplasty}
Andrabi, S.M., Sarmast, A.H., Kirmani, A.R., Bhat, A.R.: Cranioplasty:
  Indications, procedures, and outcome--an institutional experience. Surgical
  neurology international  \textbf{8} (2017)

\bibitem{chen2017computer}
Chen, X., Xu, L., Li, X., Egger, J.: Computer-aided implant design for the
  restoration of cranial defects. Scientific reports  \textbf{7}(1),  1--10
  (2017)

\bibitem{chilamkurthy2018development}
Chilamkurthy, S., Ghosh, R., Tanamala, S., Biviji, M., Campeau, N.G.,
  Venugopal, V.K., Mahajan, V., Rao, P., Warier, P.: Development and validation
  of deep learning algorithms for detection of critical findings in head ct
  scans. arXiv preprint arXiv:1803.05854  (2018)

\bibitem{Hieu2003}
Hieu, L., Bohez, E., Sloten, J.V., Phien, H., Vatcharaporn, E., Binh, P., An,
  P., Oris, P.: Design for medical rapid prototyping of cranioplasty implants.
  Rapid Prototyping Journal  \textbf{9}(3),  175--186 (Aug 2003).
  \doi{10.1108/13552540310477481},
  \url{https://doi.org/10.1108/13552540310477481}

\bibitem{jenkinson2002improved}
Jenkinson, M., Bannister, P., Brady, M., Smith, S.: Improved optimization for
  the robust and accurate linear registration and motion correction of brain
  images. Neuroimage  \textbf{17}(2),  825--841 (2002)

\bibitem{larrazabal2020}
{Larrazabal}, A.J., {Martínez}, C., {Glocker}, B., {Ferrante}, E.: Post-dae:
  Anatomically plausible segmentation via post-processing with denoising
  autoencoders. IEEE Transactions on Medical Imaging  (2020).
  \doi{10.1109/TMI.2020.3005297}

\bibitem{larrazabal2019}
Larrazabal, A.J., Martinez, C., Ferrante, E.: Anatomical priors for image
  segmentation via post-processing with denoising autoencoders. In:
  International Conference on Medical Image Computing and Computer-Assisted
  Intervention. pp. 585--593. Springer (2019)

\bibitem{lee2019tetris}
Lee, M.C.H., Petersen, K., Pawlowski, N., Glocker, B., Schaap, M.: Tetris:
  template transformer networks for image segmentation with shape priors. IEEE
  transactions on medical imaging  \textbf{38}(11),  2596--2606 (2019)

\bibitem{li2020baseline}
Li, J., Pepe, A., Gsaxner, C., von Campe, G., Egger, J.: A baseline approach
  for autoimplant: the miccai 2020 cranial implant design challenge. arXiv
  preprint arXiv:2006.12449  (2020)

\bibitem{mansilla2020learning}
Mansilla, L., Milone, D.H., Ferrante, E.: Learning deformable registration of
  medical images with anatomical constraints. Neural Networks  \textbf{124},
  269--279 (2020)

\bibitem{matzkin2020}
Matzkin, F., Newcombe, V., Stevenson, S., Khetani, A., Newman, T., Digby, R.,
  Stevens, A., Glocker, B., Ferrante, E.: Self-supervised skull reconstruction
  in brain ct images with decompressive craniectomy. MICCAI 2020  (2020)

\bibitem{monteiro2020multiclass}
Monteiro, M., Newcombe, V.F., Mathieu, F., Adatia, K., Kamnitsas, K., Ferrante,
  E., Das, T., Whitehouse, D., Rueckert, D., Menon, D.K., et~al.: Multiclass
  semantic segmentation and quantification of traumatic brain injury lesions on
  head ct using deep learning: an algorithm development and multicentre
  validation study. The Lancet Digital Health  (2020)

\bibitem{morais2019automated}
Morais, A., Egger, J., Alves, V.: Automated computer-aided design of cranial
  implants using a deep volumetric convolutional denoising autoencoder. In:
  World Conference on Information Systems and Technologies. pp. 151--160.
  Springer (2019)

\bibitem{Patravali18}
Patravali, J., Jain, S., Chilamkurthy, S.: 2d-3d fully convolutional neural
  networks for cardiac mr segmentation. In: Pop, M., Sermesant, M., Jodoin,
  P.M., Lalande, A., Zhuang, X., Yang, G., Young, A., Bernard, O. (eds.)
  Statistical Atlases and Computational Models of the Heart. ACDC and MMWHS
  Challenges. pp. 130--139. Springer International Publishing, Cham (2018)

\bibitem{shakeri2016prior}
Shakeri, M., Ferrante, E., Tsogkas, S., Lippe, S., Kadoury, S., Kokkinos, I.,
  Paragios, N.: Prior-based coregistration and cosegmentation. In:
  International Conference on Medical Image Computing and Computer-Assisted
  Intervention. pp. 529--537. Springer (2016)

\end{thebibliography}
\end{document}